\documentclass[a4paper,11pt]{article}
\usepackage{aaskaiid}
\usepackage{orcidlink}
\usepackage{xcolor}  % 颜色支持包
\usepackage{algorithm}
\usepackage{algorithmic}
\title{A Non-Negativity Iterative Approach to Image Deconvolution for SKA}
\ShortTitle{Image Deconvolution for SKA}

\author[1,2]{Le Zhang\orcidlink{0009-0009-1374-7756}}
\ShortName{Le Zhang \& Shiyu Li } % shortened name list for header 
\author[1]{Shiyu Li\orcidlink{0009-0003-5228-3487}}
\affiliation[1]{School of Physics and Astronomy, Sun Yat-sen University, 2 Daxue Road, Tangjia, Zhuhai, 519082, China}
\emailAdd{zhangle7@mail.sysu.edu.cn}
\affiliation[2]{CSST Science Center for the Guangdong-Hong Kong-Macau Greater Bay Area, Zhuhai 519082, China}
\emailAdd{lishy239@gmail.com}

\abstract{We introduce a novel algorithm for image deconvolution applicable to interferometric radio observations, based on the assumption of non-negative source fluxes. The method enables rapid and efficient image reconstruction in an iterative manner, without requiring prior knowledge or training. Its computational cost scales linearly with the number of pixels: for example, a  $512\times 512$ image can be processed in about 1-2 seconds on a standard laptop. We validate the algorithm using both point sources and an extended galaxy image, incorporating a realistic SKA-Low PSF with incomplete $uv$-coverage, though tests are conducted in noise-free simulations. Comparison with the CLEAN method demonstrates that our approach yields a good reconstruction, showing particular promise for the SKA and VLBI observations with sparse $uv$-coverage.}

\begin{document}
\maketitle

\section{Introduction}\label{sec:1}
In radio astronomy, reconstructing images from observed maps--whether from single-dish or interferometric observations--is challenging because it is an inherently ill-posed inverse problem, even in the noise-free case. A single-dish telescope forms an image by raster scanning the sky, producing an observed map that is the convolution of the true sky with the telescope beam, which blurs fine details. In interferometry, incomplete coverage of the Fourier domain leads to complex synthesized beams that also convolve with the sky image. In both cases, the observed ``dirty image'' represents the convolution of the Point Spread Function (PSF) with the true sky brightness. Reconstruction faces two main limitations: 1) the PSF’s Fourier transform provides only limited spatial frequencies, making direct inversion impossible without prior assumptions about the sky; 2) noise amplification occurs when trying to recover high frequencies through deconvolution, often destabilizing the solution.

% The current and upcoming radio telescopes, such as the Five-hundred-meter Aperture Spherical radio Telescope (FAST)~\citep{2011IJMPD..20..989N}, the Tianlai cylinder array~\citep{2012IJMPS..12..256C}, the LOw-Frequency ARray (LOFAR)~\citep{vanHaarlem:2013dsa}, the MeerKAT telescope~\citep{2016mks..confE...1J}, the Murchison Widefield Array (MWA)~\citep{Lonsdale:2009cb}, the Precision Array to Probe the Epoch of Reionisation (PAPER)~\citep{2010AJ....139.1468P}, and the Square Kilometre Array (SKA)~\citep{Weltman:2018zrl}, are poised to provide unprecedented resolution and sensitivity for exploring the Universe in intricate detail. However, they also present significant imaging analysis challenges due to the immense volume of data and high dynamic range~\citep{2009ASPC..407..375B,2013PASA...30...20N}. In this context, image deconvolution algorithms, designed to solve the inverse imaging problem, face increasing demands for accuracy, fidelity, and computational efficiency.

Deconvolution of radio point sources has long been a central challenge in radio astronomy, with the Högbom CLEAN algorithm~\citep{1974A&AS...15..417H} remaining the most widely used method for interferometric imaging. CLEAN effectively suppresses sidelobes and reconstructs complex sources but struggles with extended emission. Numerous extensions~\citep{1978A&A....65..345S,1980A&A....89..377C,1983AJ.....88..688S,2004A&A...426..747B,2008ISTSP...2..647C,2011A&A...532A..71R,2017isra.book.....T,2016Ap&SS.361..153Z} have improved its performance, yet CLEAN-restored images still depend heavily on parameter tuning. While modern packages such as CASA~\citep{2022PASP..134k4501C} and WSCLEAN~\citep{2014MNRAS.444..606O} enable faster, more automated imaging, achieving optimal results still requires extensive iterative adjustments. Alternative approaches--including compressed sensing, Bayesian imaging, and deep learning--have been widely explored in the literature, providing new perspectives and methodological viewpoints on image reconstruction.

Here we propose a novel deconvolution method that retains the densely sampled $uv$ regions--where the signal-to-noise level is high--while excluding the sparse regions that are susceptible to noise amplification due to division by small PSF Fourier amplitudes. The algorithm explicitly utilizes the well-covered PSF region and iteratively extrapolates into the uncovered $uv$ plane~\citep{mydecon}. By enforcing a non-negativity constraint at each iteration, it efficiently reconstructs the signal across the full $uv$ domain.

The paper is organized as follows: Sect.~\ref{sec:2} introduces our novel deconvolution algorithm, Sect.~\ref{sec:3} describes the simulation images and evaluates the method’s performance through three representative test cases, and Sect.~\ref{sect4} summarizes our main conclusions and discussions.

\section{Algorithm of Non-Negative Optimal-Fidelity Deconvolution}\label{sec:2}

\begin{algorithm}[ht]
\caption{Non-Negative Optimal-Fidelity Deconvolution}
\label{alg:reconstruction}
\textbf{Input:} Dirty image $x_{\rm dirty}$, PSF $h$; 
Fourier retention fraction $f_c$; 
non-negativity threshold $\epsilon$; 
smoothing scale $\sigma$; 
convergence tolerance $\tau$. \\
\textbf{Output:} Reconstructed image $x_{\rm rec}$.

\begin{enumerate}

\item Compute the Fourier transform of the PSF,
\[
H=\mathcal{F}\{h\}.
\]
Construct a binary sampling mask $M(u,v)$
by retaining the largest fraction $f_c$ of Fourier modes ranked by $|H|$.

\item Remove the mean of the dirty image and compute
\[
X_{\rm obs}=\mathcal{F}\{x_{\rm dirty}-\langle x_{\rm dirty}\rangle\}.
\]
Initialize the masked inverse-filter estimate
\[
X_{\rm model}=M\,X_{\rm obs}/H,
\]
and set $X=X_{\rm obs}$.

\item Iterate until convergence:
\[
X \leftarrow M X_{\rm model} + (1-M)X,
\]
\[
x = \mathcal{F}^{-1}\{X\}, 
\qquad x \leftarrow \max(x,\epsilon),
\]
\[
X \leftarrow \mathcal{F}\{x\}\,G,
\]
where
\[
G=\exp\!\left[-2\pi^2\sigma^2(u^2+v^2)\right]
\]
is an optional Fourier-domain Gaussian smoothing kernel.
Convergence is reached when
\[
\frac{|\sigma_k-\sigma_{k-1}|}{\sigma_{k-1}} < \tau,
\]
where $\sigma_k$ denotes the standard deviation of $x$ at iteration $k$.

\item Obtain the final reconstruction
\[
x_{\rm rec}=\mathcal{F}^{-1}\{X\},
\]
and restore the mean of  the dirty image.
\end{enumerate}
\end{algorithm}

We begin with the mathematical formulation of the observation process.  
The observed image $y$ can be expressed as
\begin{equation}
y = \mathrm{PSF} * x + n ,
\end{equation}
where $x$ is the true underlying image, $\mathrm{PSF}$ denotes the point spread function describing the instrumental response, and $n$ represents additive noise, which is assumed to follow a Gaussian distribution.  
Here, the symbol ``$*$'' denotes the convolution operation.

Unlike many conventional deconvolution algorithms that aim to minimize a regularized objective function such as
$L = \|\, y - \mathrm{PSF} * x \,\|^2 
+ \lambda_1 \|x\|_1 
+ \lambda_2 \|x\|_2^2 
+ \lambda_{\mathrm{TV}}\, \mathrm{TV}(x)$, our approach does not introduce explicit regularization terms.  
In principle, without such constraints, the deconvolution problem admits infinitely many possible solutions for $x$.  
Regularization is typically used to enforce uniqueness by globally minimizing $L$. However, the inclusion of these penalty terms inevitably shifts the solution away from the best-fit $\chi^2$ minimum.

The central ideas of our algorithm are twofold: 1) to reconstruct $x$ by focusing on the densely sampled $uv$ regions--where noise is minimal and reliable information is available--while masking out the sparsely covered regions that would otherwise cause severe noise amplification; this keeps the residual term $\|y - \mathrm{PSF} * x\|$ close to its optimal minimum, free from distortions caused by external regularization; and
2) to incorporate a physically motivated positive constraint.
The latter arises naturally from the physical property that all astrophysical sources emit positive flux, whereas negative pixel values originate purely from incomplete $uv$ coverage and the oscillatory sidelobes of the synthesized beam.
By enforcing non-negativity, the algorithm suppresses such unphysical fluctuations and implicitly performs a form of inpainting in Fourier space, progressively reconstructing the missing modes in the $uv$ plane in a self-consistent manner.
This positive prior thus serves both as a physical and mathematical regularization, stabilizing the iterative process and guiding it toward a more realistic and robust reconstruction.

Specifically, we apply a masked Fourier-domain iterative deconvolution scheme to recover the underlying sky image from the observed dirty map (see~\cite{mydecon} for details). The reconstruction workflow is summarized in Algorithm~\ref{alg:reconstruction}.

% The algorithm first removes the mean intensity of the input image to ensure that the zero-frequency component remains fixed during the iteration.
% The reconstruction then proceeds iteratively in the Fourier domain, where the observed modes are constrained by the sampling mask, and the unobserved modes are progressively refined through repeated transformations between Fourier and image space.
% In each iteration, the image is transformed back to real space to enforce a non-negativity condition, ensuring that all pixel values remain physically meaningful. A mild Gaussian smoothing is applied in the frequency domain to suppress the accumulation of high-frequency noise and stabilize convergence.
% The iteration continues until the fractional change in the image standard deviation falls below a predefined tolerance, indicating that a stable solution has been reached.
% Finally, the mean level removed at the beginning is restored to obtain the reconstructed image.
% This procedure effectively balances data fidelity and smoothness, producing a stable reconstruction even in the presence of incomplete Fourier coverage.

\section{Validation On Synthetic Data}\label{sec:3}

To validate the performance of our algorithm under controlled conditions, we conduct three test cases. We first construct a simple synthetic dataset with a map size of $128\times128$ pixels. The ground-truth image was initialized as a zero-valued map, onto which we randomly injected 100 point sources. The source positions  were uniformly distributed across the field, and their flux amplitudes were drawn from a uniform distribution in the range $[1, 100]$ (arbitrary units). This configuration represents a sparse sky model dominated by isolated compact sources, providing a clean baseline for evaluating the reconstruction fidelity of the deconvolution algorithm. 

\begin{figure*}
\centering
    \includegraphics[width=0.7\textwidth]{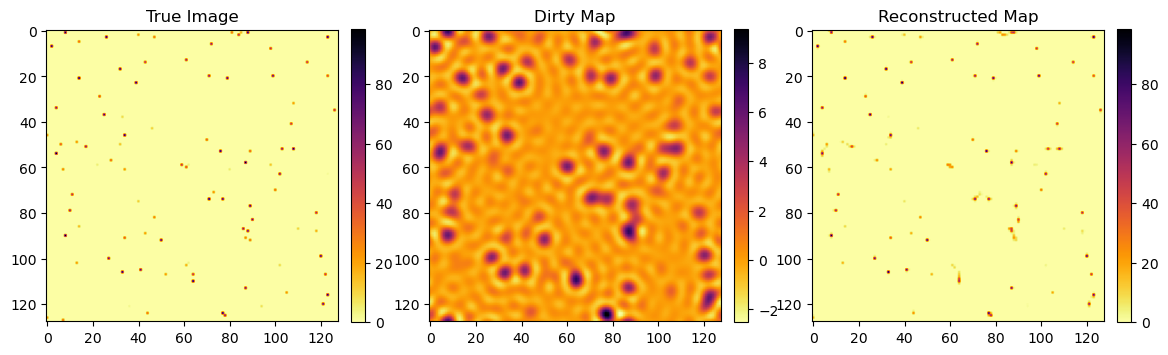}\\
    \includegraphics[width=0.7\textwidth]{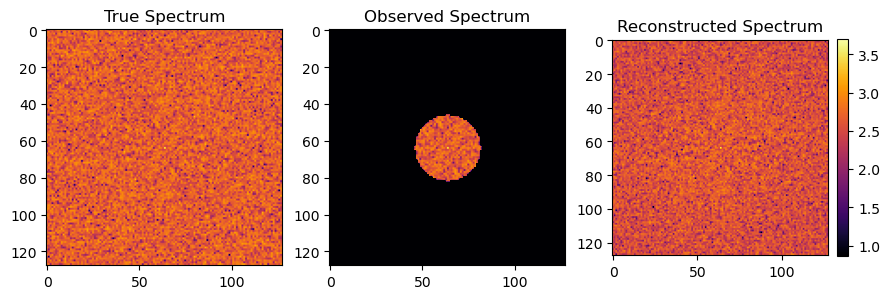}
    \caption{Comparison of the true, dirty, and reconstructed images using the point-source simulation. Note that all quantities are given in arbitrary units and are used exclusively to demonstrate the reconstruction performance. The upper panels show the images in real space, while the lower panels display the corresponding Fourier amplitude maps. A binary mask is applied in the Fourier plane to mimic incomplete $uv$ coverage. The point sources are successfully identified and reconstructed, while the ring-like artifacts arising from incomplete sampling are automatically removed by our algorithm. 
}
\label{fig:ps}
\end{figure*}

In Fig.~\ref{fig:ps}, we demonstrate the reconstruction performance of our algorithm using the point-source simulation. A binary sampling mask in the Fourier domain is applied to remove all modes with spatial frequencies above $0.2f_{\rm Ny}$, where $f_{\rm Ny}$ is the 2D Nyquist frequency (the highest spatial frequency in the image). This simulates incomplete $uv$-coverage, resulting in the characteristic ring-like artifacts observed in the dirty image. As shown, our method successfully recovers most of the point sources with high fidelity, while effectively suppressing sidelobe artifacts and eliminating the spurious ring structures caused by incomplete Fourier coverage. Meanwhile, the Fourier modes are also accurately recovered, closely matching the true spectrum.

\begin{figure*}
\centering
    \includegraphics[width=0.8\textwidth]{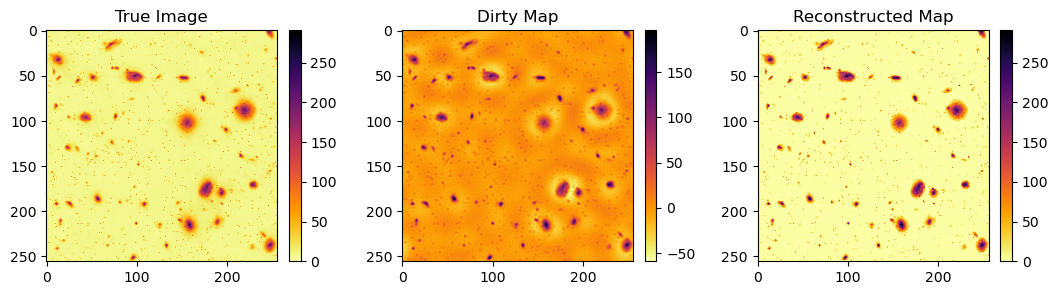}\\
    \includegraphics[width=0.8\textwidth]{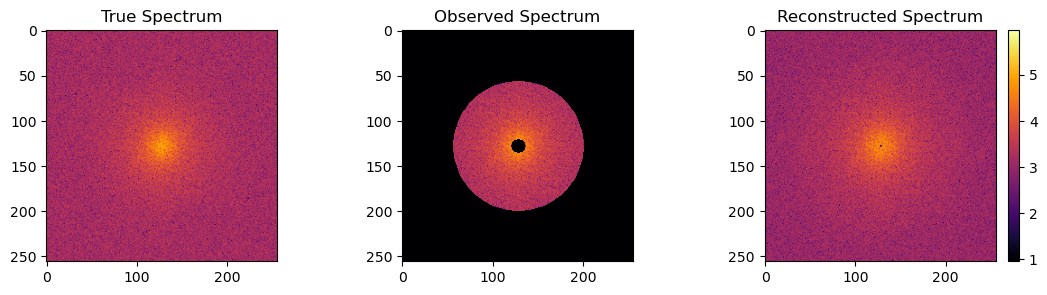}
    \caption{Similar to Fig.~\ref{fig:ps}, but for a $256\times256$ pixel simulation including both extended and point sources. The binary Fourier mask excludes the central and outer regions ($0.04<f / f_{\rm Ny}<0.4$), providing a qualitative representation of incomplete $uv$-coverage in an interferometric observation.
}
\label{fig:ps2}
\end{figure*}

We perform a similar simulation as in Fig.~\ref{fig:ps2}, but on a larger 
$256\times256$ pixel map that includes both extended and point sources. In Fourier space, a binary mask is applied that keeps only modes with normalized frequencies in the range 
$0.04<f / f_{\rm Ny}<0.4$, qualitatively mimicking the effect of incomplete $uv$-coverage in an interferometric observation.

\begin{figure*}
\centering
    \includegraphics[width=0.6\textwidth]{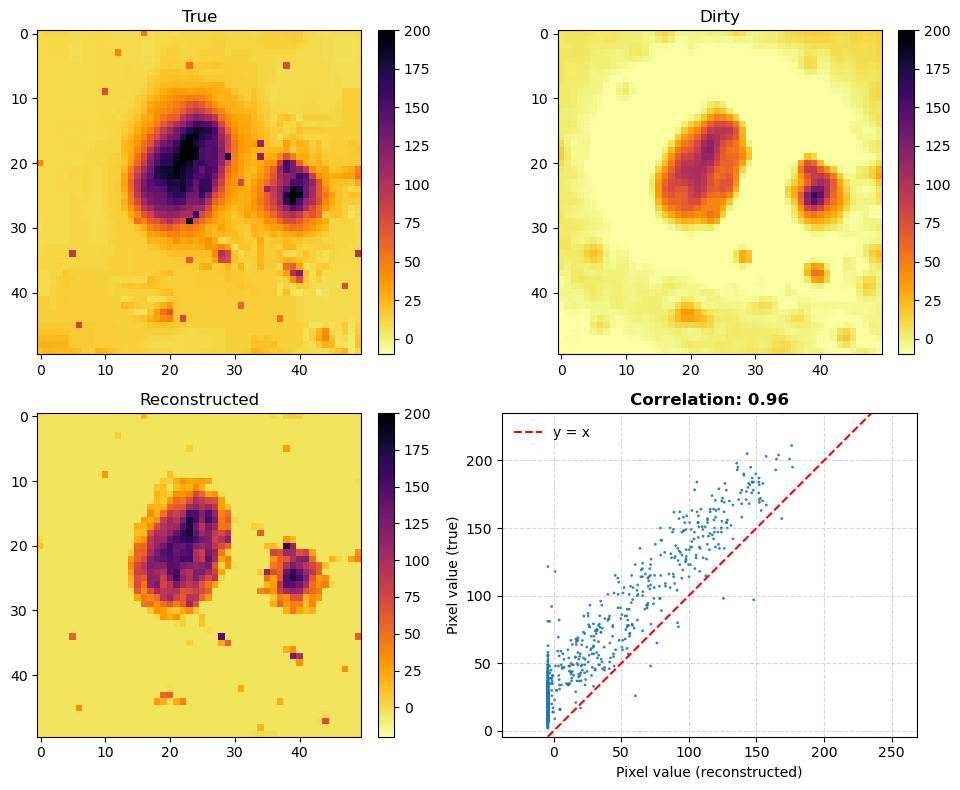}
    \caption{Zoomed-in maps from Fig.~\ref{fig:ps2}, highlighting the extended sources to clearly illustrate the deconvolution performance. The lower-right panel presents a pixel-by-pixel scatter plot, showing a correlation of $r=0.96$ across the full image.
}
\label{fig:ps3}
\end{figure*}

As seen in Fig.~\ref{fig:ps3}, the zoomed-in maps highlight the extended sources, illustrating the deconvolution performance. The reconstructed map captures both the overall morphology and fine structures with minor artifacts. The lower-right panel shows a pixel-by-pixel scatter plot, revealing a correlation of $r=0.96$ with the true image, demonstrating the accuracy of the reconstruction.

 To evaluate the reconstruction performance under realistic observational conditions, we apply a representative SKA-Low PSF to the galaxy ESO 137-001 for an image size of $512\times512$. The telescope model is based on the 512-station SKA-Low configuration in a Vogel spiral layout, ensuring uniform areal density and diverse azimuthal sampling. A 4-hour observation is simulated with 10 s integration and 100 kHz frequency resolution, and the PSF at 106 MHz is constructed with the natural weighting~\citep{Bonaldi:2025gpa}. 

\begin{figure*}
\centering
    \includegraphics[width=0.4\textwidth]{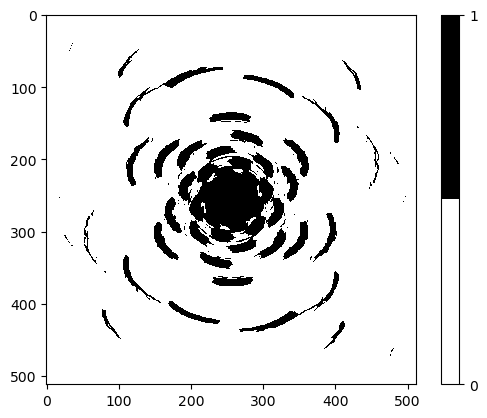}
    \caption{Binary mask in Fourier space retaining only the largest 10\% of PSF amplitude values for our fiducial SKA-Low configuration, corresponding to sparse 10\% uv coverage, typical of VLBI observations.
}
\label{fig:mask}
\end{figure*}

In our algorithm, only the largest fraction  $f_c= 10\%$ of the PSF Fourier amplitude values are retained, where $f_c$ serves as a tunable parameter controlling the effective sampling region in Fourier space. This choice is motivated by the assumption that lower-amplitude modes would be noise-dominated, although the present test cases are noise-free. Based on the Fourier amplitude of the SKA-Low PSF, we construct a binary mask to select the corresponding $uv$ points.  As seen in Fig.~\ref{fig:mask}, only the largest 10\% of Fourier amplitudes are retained, resulting in sparse 10\% $uv$ coverage, typical of VLBI-like observations. The pipeline of our algorithm typically converges within $\sim$2 s.

\begin{figure*}
\centering
  \includegraphics[width=0.9\textwidth]{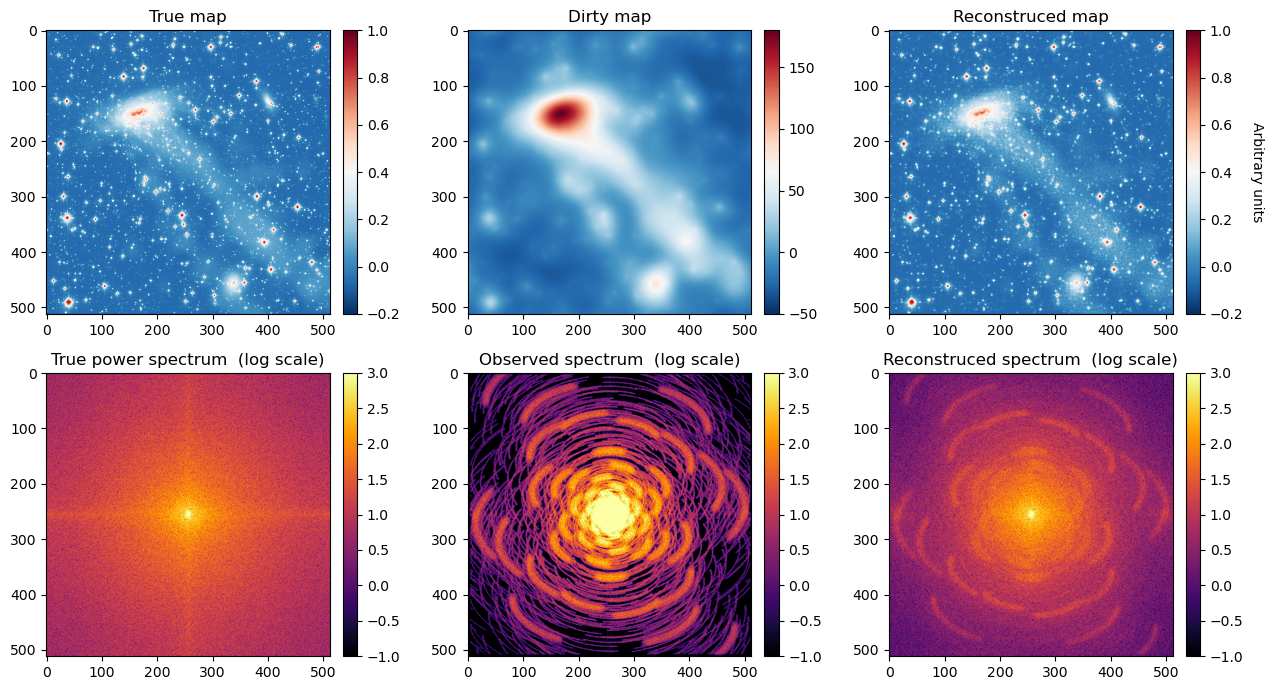}
    \caption{Illustration using a realistic SKA-Low PSF applied to the galaxy ESO 137-001 for a $512\times512$ image. The upper and lower panels present comparisons between the true, dirty, and reconstructed images in real space and Fourier space (displayed in logarithmic scale), respectively, following the same layout as Fig.~\ref{fig:ps}. Only the top 10\% of Fourier amplitudes were considered, with a binary mask applied to select $uv$ points. All values are in arbitrary units.
}
\label{fig:m101-1}
\end{figure*}

 As shown in Fig.~\ref{fig:m101-1}, the upper panels compare the true, dirty, and reconstructed images in real space, while the lower panels display their corresponding Fourier amplitudes in logarithmic scale. The reconstruction successfully suppresses PSF-induced sidelobes and restores both extended emission and compact features. In Fourier space, the recovered spectrum closely follows the true image within the sampled $uv$ coverage, while missing information in the unsampled regions is effectively inferred.

\begin{figure*}
\centering
    \includegraphics[width=0.9\textwidth]{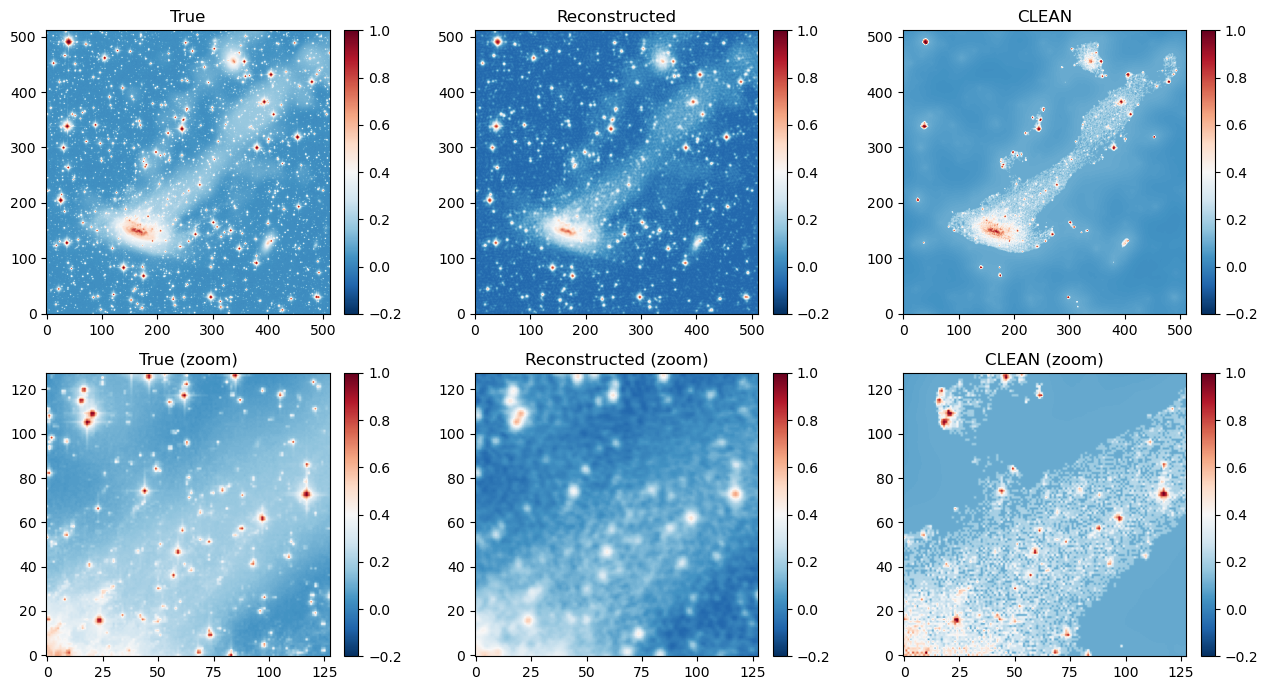}
    \caption{Similar to Fig.~\ref{fig:m101-1}, but showing a comparison between our reconstruction and the CLEAN result. The top panels present the full-image comparison, while the bottom panels show a zoomed-in region. The CLEAN deconvolution corresponds to the sum of the CLEAN ``model'' and ``residual'' image (before application of the restoring beam), obtained using a gain of 0.8 and $4\times10^{4}$ iterations. Our method more effectively recovers both diffuse and sharp structures, whereas CLEAN tends to leave extended emission and faint features partially unresolved.
}
\label{fig:m101-2}
\end{figure*}

 Fig.~\ref{fig:m101-2} further compares our reconstruction with the CLEAN deconvolution result by CASA. The CLEAN image shown here corresponds to the model component (without restoration), obtained using a loop gain of 0.8 and $5\times10^{4}$ iterations. While CLEAN is effective at identifying bright compact components, it tends to fragment diffuse emission into discrete components and leaves faint extended structures partially unresolved. In contrast, our method simultaneously preserves smooth large-scale emission and sharp morphological features, producing a reconstruction that is visually and structurally closer to the ground truth. The zoomed-in panels highlight these differences more clearly, particularly in regions dominated by low-surface-brightness emission.

\section{Enabling SKA Science Outcomes with the proposed imaging algorithm}\label{sec:4}

\label{sect4}

\begin{table}[htbp]
\centering
\caption{Comparison of reconstruction performance between this work and CLEAN.}
\begin{tabular}{lccc}
\hline
Method & Correlation $\rho$ & RRMSE & PSNR (dB) \\
\hline
This work & 0.95 & 0.26 & 31.1 \\
CLEAN     & 0.88 & 0.44 & 27.2 \\
\hline
\end{tabular}
\label{tab:recon_performance}
\end{table}

The scientific return of SKA relies on accurate imaging over a wide dynamic range, especially for faint and diffuse emission. To quantitatively evaluate the reconstruction performance, we use three standard metrics: the Pearson correlation coefficient, the relative root mean square error (RRMSE), and the peak signal-to-noise ratio (PSNR).

The Pearson correlation coefficient is
\begin{equation}
\rho = \frac{\mathrm{Cov}(x_{\mathrm{rec}}, x_{\mathrm{true}})}{\sigma_{\mathrm{rec}}\, \sigma_{\mathrm{true}}},
\end{equation}
which measures the structural similarity between the reconstructed and true images.

The relative root mean square error is
\begin{equation}
\mathrm{RRMSE} = \frac{\|x_{\mathrm{rec}} - x_{\mathrm{true}}\|_2}{\|x_{\mathrm{true}}\|_2},
\end{equation}
which quantifies the overall reconstruction error in a normalized way.

The peak signal-to-noise ratio is
\begin{equation}
\mathrm{PSNR} = 10 \log_{10} \left( \frac{\mathrm{MAX}^2}{\mathrm{MSE}} \right),
\end{equation}
where $\mathrm{MSE}$ is the mean squared error and $\mathrm{MAX}$ is the dynamic range of the true image.

Table~\ref{tab:recon_performance} shows that the proposed method achieves higher correlation, lower error, and higher PSNR than CLEAN. This indicates improved overall reconstruction fidelity. In particular, the results are consistent with improved reconstruction performance across different intensity regimes, which may lead to an enhanced effective dynamic range. This is relevant for key SKA science cases, including high-redshift galaxies, low-luminosity AGN, and diffuse synchrotron emission. 

Such improvements are also important for cosmological studies of the Epoch of Reionization (EoR), where the 21 cm signal is intrinsically weak. The improved overall correlation suggests reduced reconstruction discrepancies and fewer spurious features, which may improve the robustness of statistical analyses and foreground removal.

Overall, these results indicate that the proposed imaging algorithm has the potential to enable high-fidelity, high-dynamic-range imaging for radio observations.

% \begin{figure}[t]
% \centering
% \includegraphics[width=0.6\textwidth]{fig/6.png}
% \caption{
% \red{ Correlation between reconstructed and true sky intensity as a function of signal strength, computed in bins of true pixel intensity. The proposed method (This work) shows higher correlation than CLEAN, particularly at low intensities where CLEAN approaches zero or negative values. This suggests an improved capability in recovering faint sources and diffuse emission, which is relevant for SKA science applications.}
% }
% \label{fig:corr_vs_signal}
% \end{figure}

\section{Conclusion}

We have presented and validated an iterative deconvolution algorithm for interferometric imaging, demonstrating its effectiveness on a range of images, including the realistic galaxy ESO 137-001. By selectively reconstructing dominant Fourier modes using a SKA-Low PSF and restricting to $\sim 10\%$ $uv$ coverage, the method recovers both fine-scale structures and extended emission under sparse $uv$ sampling, with performance that is better than standard CLEAN.

This behavior suggests improved reconstruction fidelity, particularly for faint emission. The performance at low intensities is consistent with improved recovery of weak sources and diffuse structures, and may lead to an enhanced effective dynamic range. This is relevant for SKA science cases, including the detection of high-redshift star-forming galaxies, low-luminosity AGN, and large-scale diffuse synchrotron emission. The approach is also relevant for EoR studies, where the signal is intrinsically faint and sensitive to foreground residuals. The improved recovery of weak and diffuse emission suggests reduced imaging systematics, which may benefit the analysis of faint cosmological signals.

\section*{Acknowledgements}
This study is supported by the National SKA Program of China (2025SKA0160100), the National Science Foundation of China (12473097), and the Guangdong Basic and Applied Basic Research Foundation (2024A1515012309).
\bibliographystyle{abbrvnat}
\bibliography{chapter} % if your bibtex file is called example.bib

\end{document}